\documentclass[10pt,twocolumn,conference]{IEEEtran}
\usepackage{amsfonts}
\usepackage{bbm}
\usepackage{subfigure}
\usepackage{amsmath}
\usepackage{epsfig}
\usepackage{amssymb}
\usepackage{mathrsfs}
\usepackage{algorithmic}
\usepackage[ruled]{algorithm}
\usepackage{footmisc}
\usepackage{xspace}
\usepackage{graphicx}
\usepackage{array}
\usepackage{multirow}
\usepackage{picins,graphicx}

\newtheorem{theorem}{\textbf{Theorem}}

\newtheorem{definition}{\textbf{Definition}}
\newtheorem{claim}{\textbf{Claim}}

\newtheorem{remark}{\textbf{Fact}}
\newtheorem{prop}{\textbf{Property}}
\def\ie{\textit{i.e.}\xspace}
\def\etal{\textit{et al.}\xspace}

\def\iid{\textit{i.i.d.}\xspace}

\def\node{u}
\def\ngb{v}                                              
\def\charge{\lambda}                           
\def\expect{\mathbb{E}}
\def\state{\alpha}                                 
\def\bstate{\theta}                                
\def\period{T}                                        
\def\overlaptime{I}                               
\def\graph{G}
\def\set{s}                                              
\def\seta{U}                                           
\def\setb{V}                                           
\def\setc{C}
\def\setd{D}
\def\slot{\tau}
\def\edge{E}                                          
\def\prob{p}                                          
\def\probeh{\gamma}                         
\def\cat{CAT\xspace}

 \allowdisplaybreaks 

\begin{document}
%
\title{Stochastic Duty Cycling for Heterogenous Energy Harvesting Networks}
\author{
\IEEEauthorblockN{Jianhui Zhang\IEEEauthorrefmark{1}, \IEEEmembership{IEEE Member}, Mengmeng Wang, Zhi Li}\\
\IEEEauthorblockA{College of Computer Science and Technology, Hangzhou Dianzi University, Hangzhou 310018 China.\\ \IEEEauthorrefmark{1}Corresponding author, Email: jhzhang@ieee.org.}
\IEEEcompsocitemizethanks{This paper was published on the 34th IEEE International Performance Computing and Communications Conference (IPCCC 2015).}
}
\maketitle
\begin{abstract}
In recent years, there have been several kinds  of energy harvesting networks containing some tiny devices, such as ambient backscatter~\cite{liu2013ambient}, ring~\cite{gummeson2014energy} and renewable sensor networks~\cite{tang2011cool}.
During   energy harvesting, such networks suffer from the energy heterogeneity, dynamics  and prediction hardness  because the access to  natural resources  is often spatiotemporal different and timely changing among the devices.
Meanwhile, the charging efficiency is quite low especially when the power of the harvested energy is weak.
It results in the energy waste to store the harvested energy indirectly.
These features bring  challenging and interesting issues on efficient allocation of the harvested energy.
This paper studies the \emph{stochastic duty cycling} by considering  these features with the objective  characterized by maximizing the common active time.
We consider two cases: offline and online stochastic duty cycling.
For the offline case,  we design an optimal solution:  offline duty cycling algorithm.
For the online case,  we design an online duty cycling algorithm, which achieves the approximation ratio with at least $1-e^{-\probeh^2}$, where $\probeh$ is the probability able to harvest energy.
We also evaluate our algorithms with the experiment on a real energy harvesting network.
The experiment results show that the performance of the online algorithm can be very close to the offline algorithm.
\end{abstract}
\IEEEpeerreviewmaketitle

\section{Introduction}
\label{section:introduction}
The energy harvesting and communication for the tiny devices have been two interesting and challenging research directions for several years~\cite{liu2013ambient}\cite{gummeson2014energy}\cite{tang2011cool}\cite{Li2013Powering}\cite{liucof}.
Wide range of natural resources can be harvested, such as signal from TV tower~\cite{liu2013ambient}, Near Field Communication (NFC) in phone~\cite{gummeson2014energy} and solar energy~\cite{tang2011cool}.
With the harvested energy, a major task is to support the communication among devices, such as the ambient backscatter~\cite{liu2013ambient} and the renewable sensor networks~\cite{tang2011cool}.
The harvested energy is usually limited and spatiotemporal  changing because of the hardware limitation and the unpredictable variation of some environmental factors~\cite{gu2009esc}\cite{Shen2013EFCon}.
To achieve the permanent network operation, some techniques, such as duty cycling, are indispensable~\cite{tang2011cool}.

This paper focuses on the \emph{stochastic duty cycling}, which is different from previous works on the duty cycling technique.
The previous works can be roughly classified into two groups: the deterministic and stochastic duty cycling.
The deterministic duty cycling only concerns about the amount of active time in each period.
For example, some works estimate the amount of the active time for a period in advance~\cite{jiang2005perpetual}\cite{moser2010adaptive}.
In the energy harvesting networks, the deterministic duty cycling cannot follow the energy dynamic caused by the environmental factors~\cite{Buchli2014Dynamic}.
A few works focus on the stochastic duty cycling, by which both the active moments  and the amount of active time are simultaneously considered, such as for coverage~\cite{Hsin2006Randomly} and for short delay~\cite{ghidini2011energy}.
Current works on the stochastic duty cycling did not take the two factors into account together so that the harvested energy is not used with high efficiency.
The two factors are:

1) Imperfect charging efficiency. In practice, the charging efficiency of the battery or capacitor for a solar powered sensor node is often less than $75\%$~\cite{ding2000battery}, which results in the  25\% indirect energy waste to store the harvested energy.

2) Random natural energy. Some natural energy, such as solar or wind energy, is shown to be random~\cite{gu2009esc}\cite{Shen2013EFCon}\cite{ho2010markovian}, so as hard to accurately predict the profiles of the future energy for long term  because of the unpredictable variation of the natural resources, such as wind or solar power.

This paper considers the imperfect charging efficiency and the randomness of the harvested energy, and designs the stochastic duty cycling algorithms respectively for  two cases: offline and online.
The objective is to maximize the Common Active Time (\cat) over a whole period.
Under the offline case, the harvested energy over the whole period is previously known while it is not  known in advance under the online case.
The main contributions of this paper are summarized as follows:
\begin{enumerate}
  \item It is the first work, to our best knowledge, that investigates the important problem of the  stochastic duty cycling by using the harvested energy with the objective to maximize \cat.
  \item For the offline case, we  design an offline stochastic duty cycling algorithm\footnote{This paper uses the term, the offline algorithm, to represent the offline stochastic duty cycling algorithm, and  similar to the term, the online algorithm.} with the optimal solution. For the online case, we propose an online stochastic duty cycling algorithm with the approximation ratio of $1-e^{-\probeh^2}$, where $\probeh$ is the probability able to harvest energy in each time slot.
  \item  This paper has performed both theoretical analysis and experimental evaluation, and the experiment results demonstrate that the online algorithm can achieve 52.68\% \cat  of the offline algorithm on average. We analyze that the experimental result is in accordance with the theoretical result.
\end{enumerate}

\textbf{Road map.} The rest of this paper is organized as follows.
Section \ref{sec:motivation} describes the motivation and the objective of this paper. The stochastic duty cycling problem is formulated in Section~\ref{sec:problem}.
We design the optimal solution for the offline case in Section~\ref{sec:offline duty cycling} and the approximate algorithm for the online case in
Section~\ref{sec:online duty cycling}.
Section \ref{sec:related} reviews the related works on  energy harvesting  networks and  duty cycling technique.
The experiment is conducted and analyzed on the real energy harvesting networks in Section~\ref{sec:experiment}.
The whole paper is summarized in Section \ref{sec:conclusion}.

\section{Motivation}
\label{sec:motivation}
This section presents the observations  motivating us to design the new stochastic duty cycling algorithms.
One is the heterogeneity of the harvested energy because of the  spatiotemporally different access to the natural resources.
The other is the low charging efficiency because of the hardware limitation and the weakness of the harvested energy.
\subsection{Heterogenous Energy Harvesting}\label{subsect: Heterogenous harvested energy}
\begin{figure}[h]\centering
\subfigure[First day]{\label{subfig: one node one day}\includegraphics[scale=.31,bb=67 533 288 725]{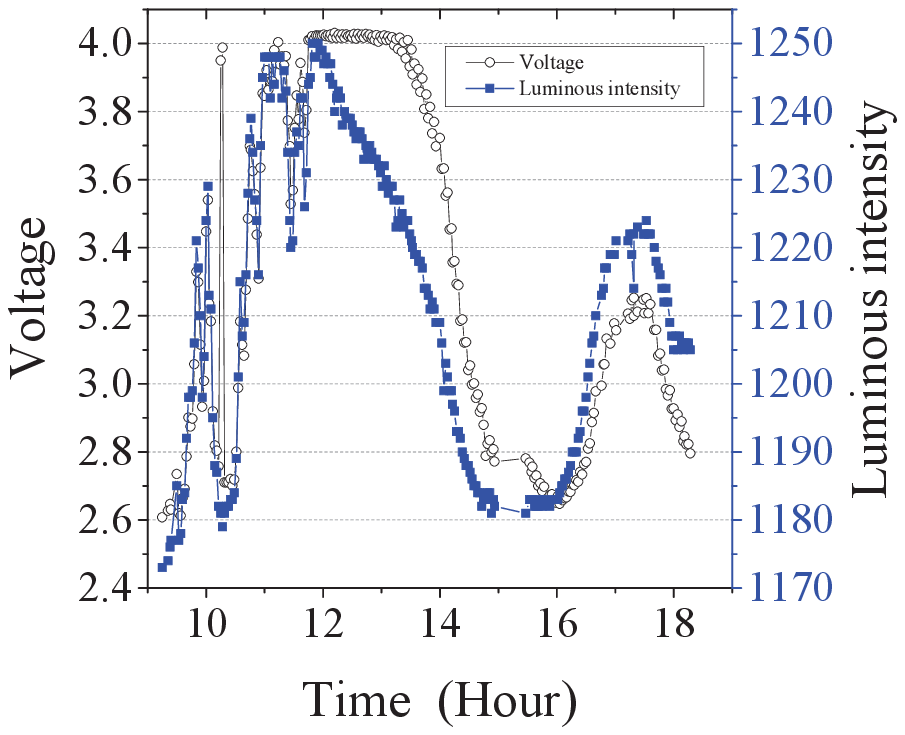}}
\hspace{0.3cm}
\subfigure[Second day]{\label{subfig:one node two day}\includegraphics[scale=.31, bb=67 533 288 725]{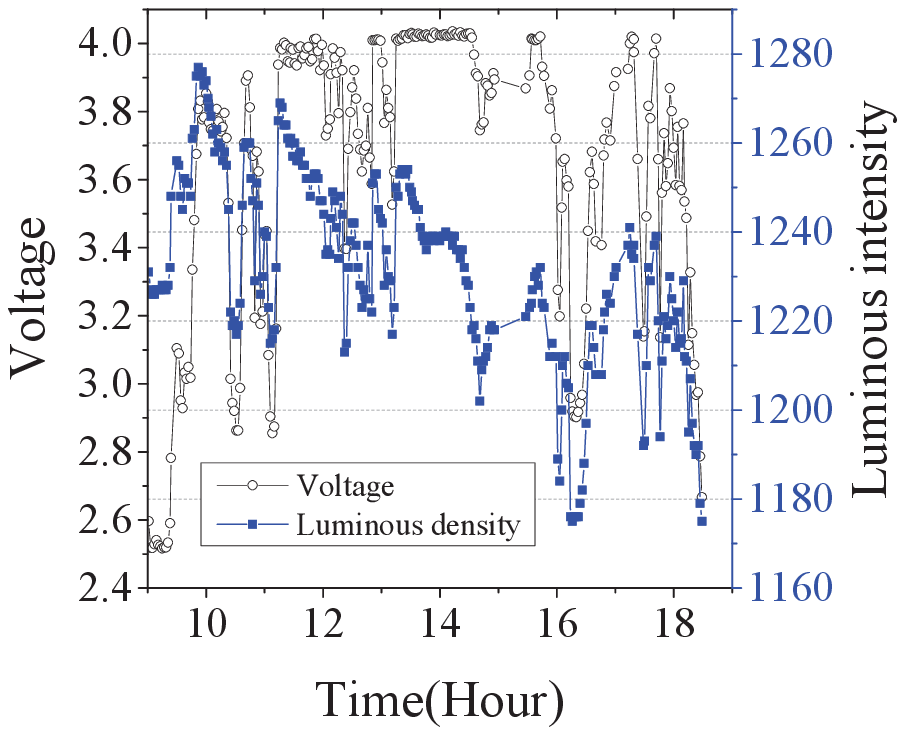}}
\hspace{0.3cm}
\subfigure[Third day]{\label{subfig:one node three day}\includegraphics[scale=.31, bb=67 533 288 725]{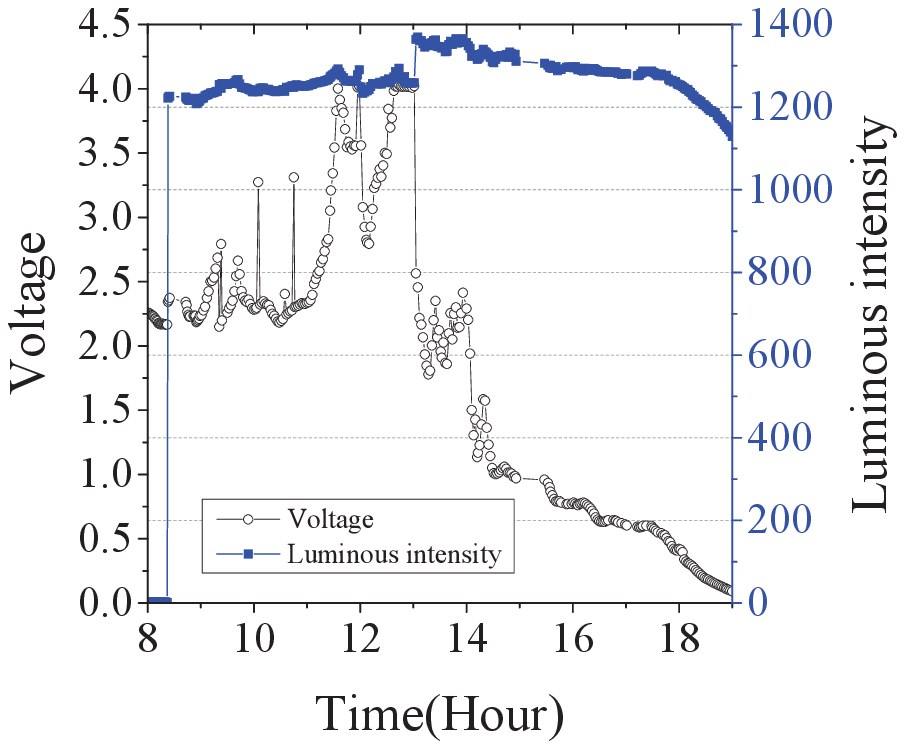}}
\caption{\label{fig:solar experiment one day} The energy harvested by a device in three days.}
\end{figure}

\begin{figure}[h]\centering
\subfigure[First device]{\label{subfig: one day node one}\includegraphics[scale=.31,bb=67 533 288 705]{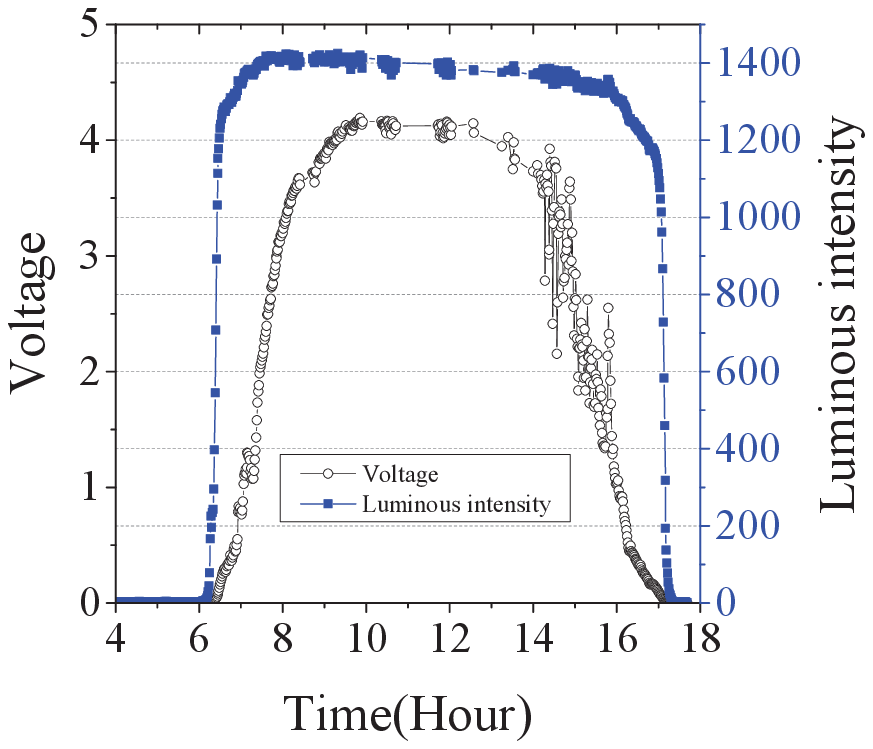}}
\hspace{0.3cm}
\subfigure[Second device]{\label{subfig:one day node two}\includegraphics[scale=.31, bb=67 533 288 705]{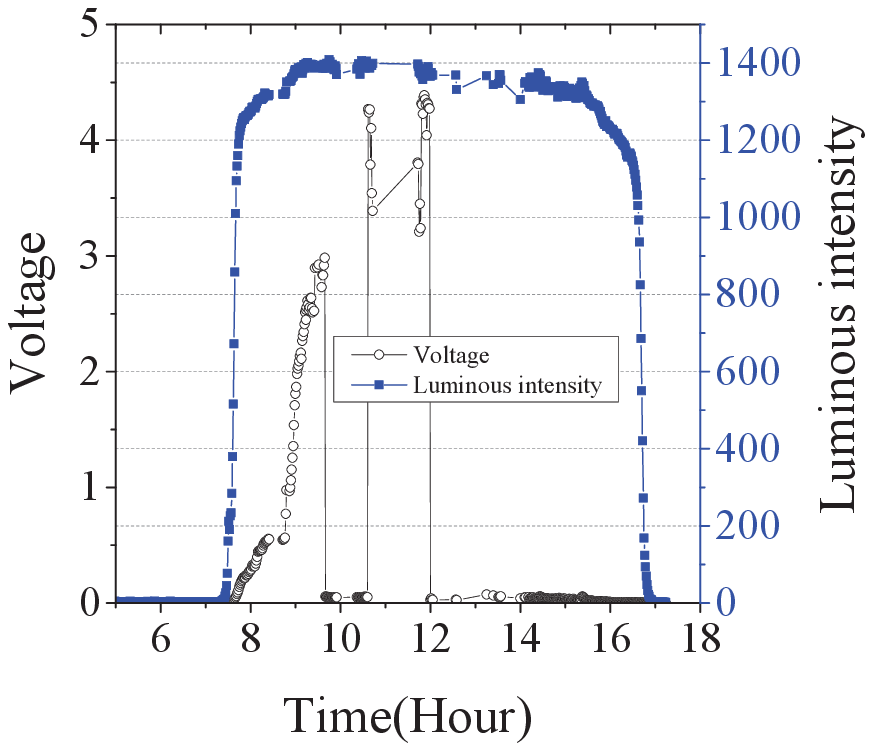}}
\hspace{0.3cm}
\subfigure[Third device]{\label{subfig:one node node three}\includegraphics[scale=.31, bb=67 533 288 705]{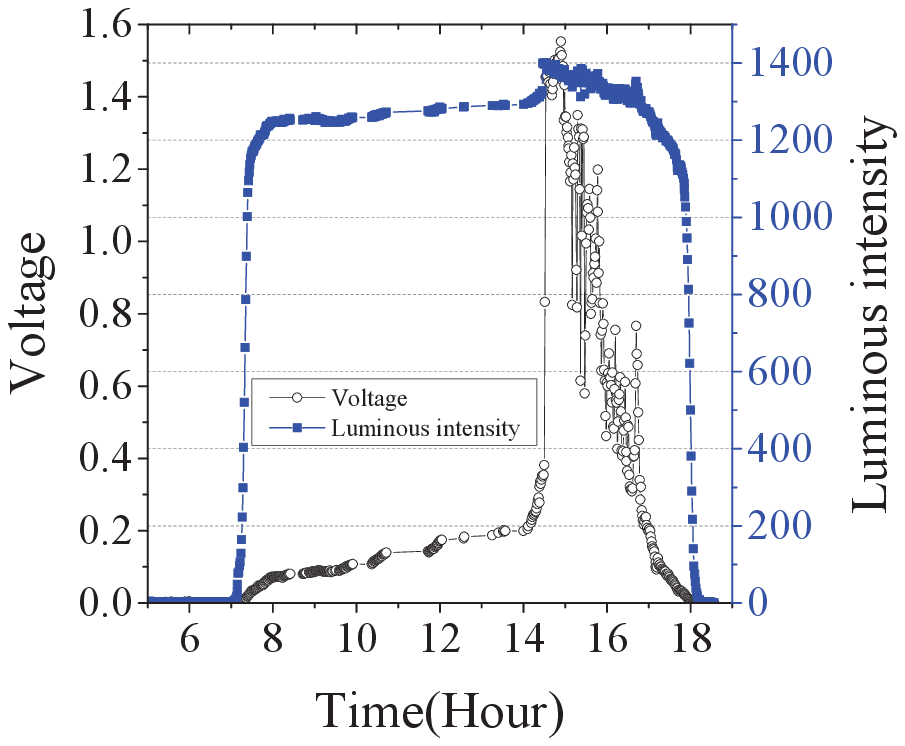}}
\caption{\label{fig:solar experiment three devices} The energy harvested by three devices in single day.}
\end{figure}

The heterogeneity  of the harvested energy appears quite frequent when the network  is deployed outdoor~\cite{gu2009esc}\cite{kansal2004performance}\cite{zhu2009leakage}.
The  unpredictable environmental factors, such as the shadow of trees, buildings or cloud, cause the device in the network to  have diverse profiles of the harvested-energy as illustrated in Figure~\ref{fig:solar experiment one day} and \ref{fig:solar experiment three devices}.
A device may have different profiles among several periods even under the similar weather conditions as illustrated in Figure~\ref{fig:solar experiment one day}.
More so, the energy profiles for several different devices vary a lot over a single period because of their different locations or some random cloud  as shown in Figure~\ref{fig:solar experiment three devices}.
As to the randomness of the harvested energy, some previous works model it as the random arrival process with \iid~\cite{Huang2013Utility} or  the Markov process~\cite{Gatzianas2010Control}\cite{he2012energy}.
\subsection{Low Charging Efficiency}
The charging efficiency can be very low because of the hardware limitation or the weak power of the harvested energy.
The popular components to store energy are battery and capacitor.
The charging efficiency of the lithium battery is about $75\%$ on average~\cite{ding2000battery}.
If the harvested energy is stored into capacitor, the energy loss is much more than that in battery because of the leakage~\cite{zhu2009leakage}.
The low charging efficiency means that  much energy is wasted inevitably if it is charged into the energy storage component.
Meanwhile, the harvested energy changes over time, and sometimes is too weak to support the normal operation of the device or to charge battery  as shown in Figure~\ref{subfig:one day node two}. In the example of this figure, the power of the harvested energy is lower than 3 V before 10 o'clock so as not to charge the battery.

The devices can only communicate with each other during their \cat, such as the data transmission among the nodes in ad hoc and wireless sensor networks.
To improve the  \cat under the low charging efficiency, it is better to use the harvested energy directly rather than to store it firstly.
The  heterogeneity of the harvested energy constrains that the devices cannot use the harvested energy directly and synchronously in every moment.
Furthermore, the harvested energy arrives randomly, which makes it quite challenging to arrange the \cat among devices.
These observations motivate  us to design new stochastic duty cycling algorithms in order to improve the utilization efficiency of the harvested energy especially when the harvested energy is heterogenous and the charging efficiency is low.
\subsection{Objective}\label{subsec:object}
Suppose that the period $\period$ is divided into several time slots $\slot=1,\cdots,|\period|$. The metric, \cat, is defined by the following statement:
\begin{definition}[Common Active Time (\cat)]\label{def:overlap time}
Suppose the active time of the devices $\node$ and $\ngb$ are respectively the time slot sets $\set$ and  $\set'$, $\set,\set'\subset\period$. Their \cat over the period $\period$ $\overlaptime_{\node\ngb}(\period)$ is the intersection of the two sets, \ie $\set\cap\set'$.
\end{definition}

Next, we design a way to determine the \cat so that we can measure the impact of both of the heterogenous harvested energy and the low charging efficiency.
Let $\state$ denote the decision of the device.
If  $\node$ is active,  $\state_\node=1$ and 0 otherwise.
Let $\bstate$ denote the energy state.  If the device  $\node$ can access the natural resources, $\bstate_\node=1$, and 0 otherwise.
The \cat between the device $\node$ and its neighbor $\ngb$ can be calculated by the following facts.
If $\bstate_\node(\slot)=1$ in time slot $\slot$,  then $\node$ can use the energy directly in this slot $\slot$ or store it with the charging efficiency $\charge$.
When $\bstate_\node(\slot)=1$ and $\bstate_\ngb(\slot)=1$,  $\state_\node(\slot)=1$ and $\state_\ngb(\slot)=1$, the \cat is 1, \ie $\overlaptime_{\node\ngb}(\slot)=1$.
When $\state_\node(\slot)=0$ and $\state_\ngb(\slot)=0$, $\overlaptime_{\node\ngb}(\slot)=0$ and the device can store the harvested energy if available.
When $\state_\node(\slot)=1$ and $\state_\ngb(\slot)=1$ and not both $\node$ and $\ngb$ can access the natural energy simultaneously, one or both of  $\node$ and $\ngb$ have to use their stored energy.
In this case, $\overlaptime_{\node\ngb}(\slot)=\charge$.
Therefore, the \cat in each time slot $\slot$ can be calculated by the following equation:
\begin{eqnarray}\label{equ:overlap time}
\overlaptime_{\node\ngb}=
\left\{ \begin{aligned}
1,\hspace{1.5cm} \state_\node=\state_\ngb=1\ \&\ \bstate_\node=\bstate_\ngb=1\\
0,\hspace{3.1cm} \state_\node=0\ or\ \state_\ngb=0\\
\charge,  \hspace{0.5cm} \state_\node =\state_\ngb=1\ \&\  \bstate_\ngb = 0\neq\bstate_ \node= 0
\end{aligned} \right.
\end{eqnarray}

The goal of this paper is  to maximize the overall \cat between the device and its neighbor over the period $\period$.
To achieve the goal, the device must  determine a sequence of decision $\state(\slot)$ over $\period$ so that the overall \cat can be maximized, \ie $\max\Sigma^\period_{\slot=1}\overlaptime_{\node\ngb}(\slot)$.

Most symbols used in this paper are summarized in Table~\ref{table:notations}.
\begin{table}[h]
\caption{Symbol and meaning\label{table:notations}}
\vspace*{-12pt}
\begin{center}
\def\temptablewidth{0.45\textwidth}
{\rule{\temptablewidth}{1pt}}
\begin{tabular*}{\temptablewidth}{@{\extracolsep{\fill}}c|l|c|lr}\centering
Sym. & Description &Sym. & Description \\ \hline
$\period$&Period & $\slot$& Slot \\
$\node$& Device&$\ngb$&Device neighbor\\
$\seta$&Device set&$\setb$&Neighbor set\\
$\edge$&Edge set&$\graph(\seta,\setb,\edge)$&Graph\\
$\charge$&Charging efficiency&$\overlaptime$&\cat\\
$\bstate$&Energy state&$\probeh$& Probability of $\bstate=1$\\
  \end{tabular*}
{\rule{\temptablewidth}{1pt}}
\end{center}
\end{table}
\section{Preliminaries}
\label{sec:problem}
This section gives the model of the energy harvesting process and network, and then formulates the stochastic duty cycling problem.
\subsection{Model}\label{subsect:network model}
In the energy harvesting networks, the device is able to harvest natural energy and to communicate with its neighbor.
Each device can store  the harvested energy with the charging efficiency $\charge$ or use it directly.
Because of the hardware limitation, the device can harvest tiny energy and thus cannot support its normal operation and charge its battery with the harvested energy simultaneously.
The harvested energy can support the device's normal operation  or can charge the battery if and only if its power is over a threshold.
If the  harvested energy is higher than the threshold, $\bstate=1$ and 0 otherwise.
Assume that the energy that the device is able to harvest is one unit through a single time slot.
As previous works, we model the energy harvesting process as the random arrival process with \iid over time slots, and thus $\bstate=1$ with probability $\probeh$ as previous works~\cite{Huang2013Utility}\cite{Gatzianas2010Control}.
This paper considers the point-to-point (P2P) communication model, \ie one device communicates with only one another device in a period.
This kind of model is very popular in many applications, such as data collection in sensor networks.
\subsection{Stochastic Duty Cycling Problem}\label{subsect:problem formulation}
Each device need determine its decision for each time slot to allocate its energy under the constraints of the heterogeneity of the harvested energy  and the low charging efficiency when the accesses to harvest energy arrives \iid.
With the object in Section~\ref{subsec:object}, the \emph{stochastic duty cycling problem} in this paper is to maximize the \cat over the period $\period$. With Equation (\ref{equ:overlap time}), the stochastic duty cycling can be formulated as the following programming.
\begin{align}\label{equ:max overlap}
&\max\sum\limits_{\slot\in\period}\overlaptime(\slot)  \\
& \text{s.t.}\hspace{0.3cm} \sum\limits_{t=1}^\slot\state(t)\leq\sum\limits_{t=1}^\slot\bstate(t),  \hspace{0.3cm}\forall \text{each device} \\
& \hspace{0.7cm} \state(\slot),\bstate(\slot)\in\{0,1\}, \hspace{0.3cm}\forall \slot\in\period
\end{align}
where the first constraint means that the consumed energy should be less than the harvested.

This paper considers the stochastic duty cycling in two cases:  offline and online.
\begin{definition}[Offline Stochastic Duty Cycling]
The device has the complete knowledge of the harvested energy of its own and neighbor over the whole period, \ie $\bstate(\slot)$, $\slot\in\period$. They then determine the decision sequence $\state(\slot)$, $\slot\in\period$.
\end{definition}

The offline case has a few applications in reality, for example, the state of the harvested-energy in current period can be estimated precisely when the period has similar condition to access the natural energy with the previous periods~\cite{Piorno2009Prediction}.
This paper considers this case as a baseline for comparison with the online case.

\begin{definition}[Online Stochastic Duty Cycling]
In each time slot $\slot$, the device determines the decision $\state(\slot)$ while it has no knowledge of  future harvested energy of its own and neighbor.
\end{definition}

This case is more practical and applicable to the real scenarios of the energy harvested networks, where the harvested energy is heterogenous and stochastic.
\section{Offline Stochastic Duty Cycling}
\label{sec:offline duty cycling}
In this section, we present the overview of the offline  stochastic duty cycling algorithm, and then give the algorithm details and analyze  its theoretical performance finally.
\subsection{Overview}
This section designs the offline algorithm for the offline case by two steps.
Since under the offline case, the device has the complete knowledge of the harvested energy of its own $\node$ and neighbor $\ngb$, \ie $\bstate_\node(\slot)$ and $\bstate_\ngb(\slot)$, $\forall \slot=1,\cdots,\period$, we   construct an \emph{energy state graph} $\graph(\seta,\setb,\edge)$ to describe the knowledge of the device and its neighbor in the first step.
$\seta$ and $\setb$ are the sets of vertexes, which represent the time slots with the energy state $\bstate=1$.
$\edge$ is the set of the edges among the vertexes respectively from  the sets $\seta$ and $\setb$.
The graph then can be mapped to a bipartite graph, and is used to  characterize the timing and intercover relations between the harvested energy states of neighboring devices.
In the second step, we transfer the energy state graph to the \emph{offline duty cycling graph}.
With this graph, we are able to convert the offline duty cycling problem to the maximal weighted matching problem in the bipartite graph.
Theoretical proof shows that the offline duty cycling graph is the maximum weighted   matching between the two sets $\seta$ and $\setb$, and accordingly the maximal \cat over the period.
\subsection{Offline Duty Cycling Algorithm}\label{subsec:offline duty cycling}
\begin{figure}[h]\centering
  \includegraphics[scale=1,bb=156 322 346 371]{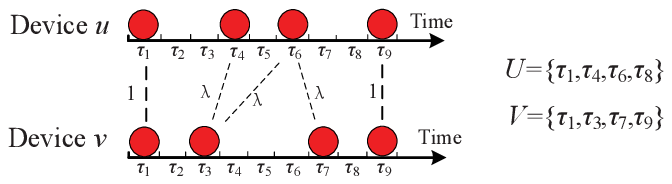}
  \caption{\label{fig:energy state graph} Energy state graph $\graph(\seta,\setb,\edge)$.}
\end{figure}

We firstly present the properties of the heterogeneous energy harvesting, and then describe the construction of the energy state graph.
For statement convenience, this paper defines some terms.
Let \emph{vertex} denote the time slot, at which the state of the harvested energy is $\bstate=1$ as the red circle in Figure~\ref{fig:energy state graph}.
For example, $\slot_1(\node)$ is the vertex of $\node$ at time slot $\slot_1$ in Figure~\ref{fig:energy state graph} when $\bstate_u(\slot_1)=1$.
If two vertexes respectively from the two vertex sets $\seta$ and $\setb$ appear at the same time slot, such as $\slot_1(\node)$ and $\slot_1(\ngb)$, $\node\in\seta$ and $\ngb\in\setb$, they are called \emph{synchronous vertexes}.
Otherwise, they are called  \emph{asynchronous vertexes}.
The edge connecting synchronous vertexes is called \emph{synchronous edge}, and otherwise, is called \emph{asynchronous edge}.
The device can use the harvested energy to communicate with its neighbor at any subsequent time slot.
Each vertex has some optional neighboring vertexes to connect, and thus has several possible edges as the dash lines  shown in Figure~\ref{fig:energy state graph}.

Notice that the energy harvested in one time slot can support the device's communication for at most one slot.
Accordingly, each vertex can have only one deterministic edge. We have the following property:
\begin{prop}\label{prop: one deterministic edge}
There is at most one deterministic edge incident to each vertex.
\end{prop}

Let the \cat represent the weight of the edge, and recall the definition of the \cat in Definition~\ref{def:overlap time}.
We have the following property:
\begin{prop}\label{prop: weight}
The synchronous edges have the same \cat, and the same to the asynchronous edges. The synchronous edge have more \cat than the asynchronous edge with the energy harvested in one time slot.
\end{prop}

We can assign the synchronous edge with weight one and asynchronous edge with weight $\charge$ according to Property \ref{prop: weight}.
The example in Figure~\ref{fig:energy state graph} illustrates the above properties.
The synchronous edges $(\slot_1(\node),\slot_1(\ngb))$ and  $(\slot_9(\node),\slot_9(\ngb))$ are assigned the weight one.
The asynchronous edges $(\slot_4(\node),\slot_3(\ngb))$, $(\slot_6(\node),\slot_3(\ngb))$ and $(\slot_6(\node),\slot_7(\ngb))$ are assigned the weight $\charge$.

The method to construct the offline duty cycling graph is given by the following steps:
\begin{enumerate}
  \item Construct the edges between the pairs of the synchronous vertexes.
  \item Each remaining vertex  searches backward and connects to the nearest vertex of its neighbor device.
\end{enumerate}

\begin{figure}[h]\centering
\includegraphics[scale=1,bb=203 431 390 478]{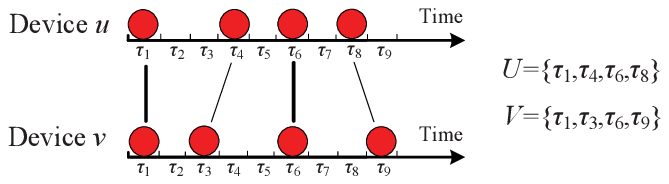}
\caption{\label{fig:reduced energy state graph} Offline duty cycling graph $\graph(\seta,\setb,\edge')$.}
\end{figure}

The example in Figure~\ref{fig:reduced energy state graph} illustrates the way to construct the offline duty cycling graph.
In the first step,  the deterministic synchronous edges: $(\slot_1(\node),\slot_1(\ngb))$ and  $(\slot_6(\node),\slot_6(\ngb))$, are found one by one according to Property~\ref{prop: one deterministic edge}, \ie making the decision $\state_\node(\slot_1)=\state_\ngb(\slot_1)=1$ and $\state_\node(\slot_6)=\state_\ngb(\slot_6)=1$.
In the second step, the deterministic asynchronous edges are found.
The vertex $\slot_3(\ngb)$ searches another vertex in the set $\seta$ to establish a deterministic asynchronous edge.
$\slot_3(\ngb)$ can only search backward, \ie the vertex in $\seta$, which appears later than $\slot_3(\ngb)$, since the device can use the energy only after it is harvested and stored.
Furthermore, $\slot_3(\ngb)$ need only search the nearest vertex $\slot_4(\node)$ in $\seta$ since all asynchronous edges have the same weight $\charge$. Then the asynchronous edge $(\slot_4(\node),\slot_3(\ngb))$ is found, \ie make decision $\state_\node(\slot_4)=\state_\ngb(\slot_4)=1$.
Similarly, the deterministic edge $(\slot_8(\node),\slot_9(\ngb))$ is established, \ie make the decision $\state_\node(\slot_9)=\state_\ngb(\slot_9)=1$.

 Recall that the goal of the offline algorithm is to assign the harvested energy to the proper time slots so that the \cat between the device $\node$ and its neighbor $\ngb$ is maximized.
From the energy state graph, we can observe that each edge represents a possible assignment of the harvested energy and the edge weight is the corresponding \cat by the energy assignment.
We thus have the following claim.
\begin{claim}
The offline duty cycling algorithm that assigns the harvested energy to maximize the \cat corresponds to a solution to finding the maximum weighted matching between the vertexes of the energy state graph.
\end{claim}

To maximize the overall \cat,  we next design the offline duty cycling algorithm to construct the offline stochastic duty cycling graph.
According to Property~\ref{prop: weight}, the edge with weight one must be first  selected.
According to Property~\ref{prop: one deterministic edge},  each vertex can match only one another vertex.
With the offline duty cycling graph, we  design the offline duty cycling algorithm and summarize it in Algorithm~\ref{alg:offline duty cycling algorithm}.
\begin{algorithm}[h]
\caption{Offline Duty Cycling Algorithm}
\label{alg:offline duty cycling algorithm}
\textbf{Input:} The harvested energy  states: $\bstate(\slot)$, $\slot=1,\cdots,\period$.\\
\qquad \textbf{Output:} The edge set $\edge'$, \ie, a weighted matching among the energy state graph.\\
\begin{algorithmic}[1]
\STATE \label{alg:step 0}Represent the time slot of $\bstate(\slot)=1$ with the vertex;
\STATE \label{alg:step 1}Construct an edge between each pair of synchronous vertexes and assign it with the weight 1;
\STATE \label{alg:step 2}In the remaining vertexes, each vertex $\slot(\node)$ searches backward for the nearest vertex $\slot(\ngb)$ and establishes an asynchronous edge with weight $\charge$;
\STATE \label{alg:step 3}In the remaining vertexes, each vertex $\slot(\ngb)$ searches backward for the nearest vertex $\slot(\node)$ and establishes an asynchronous edge with weight $\charge$;
\end{algorithmic}
\end{algorithm}
\subsection{Theoretical Performance}
By Algorithm~\ref{alg:offline duty cycling algorithm}, we can construct the offline duty cycling graph $\graph(\seta,\setb,\edge')$, which connects the vertexes of neighboring devices with the weighted deterministic edges.
 The theorem~\ref{theorem: maximal matching} shows that the edge set $\edge'$ is the maximal weighted matching between the two sets: $\seta$ and $\setb$, \ie the maximal \cat over the period.
\begin{theorem}[Optimality]\label{theorem: maximal matching}
The edge set $\edge'$ of the offline duty cycling graph is the maximum weighted bipartite matching between the two sets $\seta$ and $\setb$.
\end{theorem}
\begin{IEEEproof}
We prove by contradiction.
Suppose that there is another set $\edge''$ having the overall weight higher than $\edge'$.
Since the edges have two kind of weights: 1 and $\charge$, we divide the edge set into two subsets: one $\edge_1$ with weight 1, and  another $\edge_\charge$ with weight $\charge$.
We have $\edge'=\edge'_1\cup\edge'_\charge$ and $\edge''=\edge''_1\cup\edge''_\charge$.
There must be $|\edge''_1|>|\edge'_1|$ or $|\edge''_\charge|>|\edge'_\charge|$, where $|\cdot|$ is the set cardinality.
If $|\edge''_1|>|\edge'_1|$, there is at least one another synchronous edge, which is not included in $\edge'_1$. But it contradicts Algorithm~\ref{alg:offline duty cycling algorithm} since the algorithm searches all synchronous edges firstly in the step~\ref{alg:step 1}.
Similarly, we can prove the case, in which $|\edge''_\charge|>|\edge'_\charge|$, also contradicts Algorithm~\ref{alg:offline duty cycling algorithm} since the algorithm searches all asynchronous edges in the step~\ref{alg:step 2} and \ref{alg:step 3}.
\end{IEEEproof}

This above theorem means that the maximal \cat over the period can be maximized by Algorithm~\ref{alg:offline duty cycling algorithm}.
The maximal \cat over the period is given by the following theorem.
\begin{theorem}[Expected \cat]
The expected \cat over the period $\period$ by Algorithm~\ref{alg:offline duty cycling algorithm} is $|\period|\probeh[\probeh+2\charge(1-\probeh)]$ per  device.
\end{theorem}
\begin{IEEEproof}
Recall that the harvested energy state is \iid over time slots in Section~\ref{subsect:network model}.
The device can harvest energy in each time slot with $\probeh$.
The expected number of synchronous edges over the period is $\sum_{\slot_i=1}^\period\probeh^2$.
The expected number of asynchronous edges over the period is $\sum_{\slot_i=1}^\period 2\probeh(1-\probeh)$.
Thus the expected overall \cat is
$\expect(\overlaptime(\period))=|\period|\probeh[\probeh+2\charge(1-\probeh)]$
\end{IEEEproof}

There are some existing algorithm to find the maximal weighted  bipartite matching~\cite{Madry2013Navigating}. Their complexity is quite high.
In this paper, the time complexity of Algorithm~\ref{alg:offline duty cycling algorithm}  is not as high as the previous works.
In the step~\ref{alg:step 0} of Algorithm~\ref{alg:offline duty cycling algorithm}, each device costs $|\period|$ steps to identify the harvested energy state of each time slot.
In the step~\ref{alg:step 1}, it takes at most $\max\{|\seta|,|\setb|\}$ steps to establish synchronous edges among the vertexes.
In the step~\ref{alg:step 1}, it takes at most $|\seta||\setb|$ steps to establish asynchronous edges among the vertexes.
Therefore, the overall time complexity is $\max\{|\period|, |\seta||\setb|\}$ since $|\seta|$ and $|\setb|$ are less than $|\period|$.
\begin{theorem}[Time complexity]
The time complexity of Algorithm~\ref{alg:offline duty cycling algorithm} is $O(\max\{|\period|, |\seta||\setb|\}\})$.
\end{theorem}
\section{Online Stochastic Duty Cycling}
\label{sec:online duty cycling}
In many real applications, the networks cannot know the profiles of the harvested energy in advance.
This section considers the online stochastic duty cycling.
We firstly describe the online stochastic duty cycling problem and then give its mathematical formulation, and finally present an approximate algorithm for this problem.
Its theoretical performance is analyzed in the last part of this section.
\subsection{Problem Description}
Recall that the natural resources is modeled as the random arrival process.
Under the online duty cycling, the device and its neighbor have no future information of the profiles of their harvested energy in advance rather than the historical information of their harvested energy.
By the historical  information,  the device need make online decision to be active, or to store the harvested energy in order to maximize the overall \cat.
If the device can harvest energy in this slot, it can be active by using the harvested energy directly. Otherwise, it has to use the stored energy or to sleep.
When it decides to sleep, it can store the harvested energy if the energy state $\bstate=1$.
The device can use the harvested energy to establish an asynchronous edge with its neighbor if the neighbor has the stored energy.
To use the natural energy with high efficiency,  the device prefers to be active. But it may lose the access to store the harvested energy if its neighbor cannot harvest energy simultaneously.
The online stochastic duty cycling problem is how the device makes online decision: sleep or active, in each time slot so that the overall \cat can be maximized.
\subsection{Algorithm Design}\label{subsec:algorithm design}
\begin{figure}[h]\centering
\includegraphics[scale=1,bb=201 393 392 440]{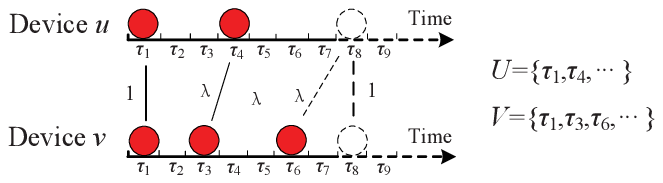}
\caption{\label{fig:online energy state graph} Online duty cycling graph $\graph'(\seta,\setb,\edge)$.}
\end{figure}

This section presents the detailed design of the online  algorithm.
To illustrate the algorithm clearly, this  section constructs an online stochastic duty cycling graph to describe the online duty cycling process as the example in Figure~\ref{fig:online energy state graph}.
In this figure,  the current time slot is $\slot_7$, and the device has to make decision: sleep or active, for next time slot $\slot_8$.
Since the harvested energy is random arrival under the online case, the device may harvest the energy in next time slot $\slot_8$ with some probability, and the same to its neighbor.
Let the dashed circle denote the possible vertexes as the vertexes $\slot_8(\node)$ and $\slot_8(\ngb)$ in Figure~\ref{fig:online energy state graph}.  Accordingly, the possible vertexes can establish possible edges with other vertexes.
With the  graph,  the online duty cycling problem is equivalent to finding the maximal matching between the vertexes of the device and its neighbor when the vertexes arrive online.

The idea behind the online duty cycling algorithm is to capture the  synchronous edge as many as possible so as to improve the overall \cat as much as possible.
The online duty cycling algorithm works as follows.
At each coming time slot, the possible vertexes $\slot(\node)\in\seta$ or  $\slot(\ngb)\in\setb$ arrive together with the possible edges. The vertexes arrive online  and in random order.
The algorithm always has to either assign the current vertex in $\seta$ to connect one of the unconnected vertexes of its neighbor in $\setb$, \ie select synchronous edges for the newly arrived vertexes under the constraint given by Property~\ref{prop: one deterministic edge}.
Our algorithm selects the synchronous and asynchronous edges with certain probability, $\probeh$, which is the statics of the  historical information of the harvested energy in the probability. According to the assumption in Section~\ref{subsect:network model}, we need not to calculate the probability.
Thus, each device makes decision online with the probability $\probeh$ to be active in each time slot, \ie $\state(\slot)=1$, and $\state(\slot)=0$ otherwise.

We summarize the online duty cycling algorithm as following pseudocode in Algorithm~\ref{alg:online duty cycling algorithm}.
\begin{algorithm}[hptb]
\caption{Online Duty Cycling Algorithm}
\label{alg:online duty cycling algorithm}
\textbf{Input:} The probability $\probeh$ and the period $\period$.\\
\qquad \textbf{Output:} The edge set $\edge$.\\
\begin{algorithmic}[1]
\FOR{$\slot=1,\cdots,\period$}
\STATE Each device makes decision to be active with probability $\probeh$ and to sleep with probability $1-\probeh$;
\STATE If the active decision is chosen and the vertexes $\slot(\node)$ and $\slot(\ngb)$ arrive, \ie,  adding the edge $(\slot(\node),\slot(\ngb))$ to $\edge$;
\STATE If the sleep decision is chosen and the vertex $\slot(\node)$ arrives actually, the vertex connects to one unconnected vertex $\slot'(\ngb)$, where $\slot'<\slot$, and then add the edge $(\slot(\node),\slot'(\ngb))$ to $\edge$.  Otherwise, leave the vertex $\slot(\node)$ unmatched.
\ENDFOR
\end{algorithmic}
\end{algorithm}
\subsection{Theoretical Performance}\label{subsec:algorithm performance}
This section presents the theoretical performance of the  algorithm~\ref{alg:online duty cycling algorithm} by approximation analysis.
An algorithm for a maximization problem is said to be $\rho$-\emph{approximate algorithm} if the \cat obtained by the algorithm is at least $1/\rho$ of the optimal algorithm.

We characterize one useful extension of the balls-in-bins problem, where this paper is interested in the distribution of certain functions of the bins by using the Azuma's inequality on appropriately defined Doob's Martingales.
The extension is given in the following fact:
\begin{remark}\label{fact: ball bin}
Suppose that there are $n$ balls and $n$ bins out of $m$. The balls are thrown into bins, \iid, with uniform probability over the bins.
Let $A$ be a certain subset of bins, and $S$ be a random variable that equals the number of bins from $A$ with at least one ball.
With probability at least $1-2e^{-\varepsilon^2m/2}$, for any $\varepsilon>0$, we have:
 \begin{equation}
|A|[1-e^{-\frac{n}{m}}]-\varepsilon m\leq\expect(S)
 \end{equation}
\end{remark}
\begin{IEEEproof}
The probability that each bin has at least one ball is $1-(1-\frac{1}{m})^n$ so the expectation of $S$ is $\expect(S)=\sum_{a\in A}[1-(1-\frac{1}{m})^n]$. By the standard identities, we have:
 \begin{equation}\label{equ:ball in bin}
\sum_{a\in A}[1-e^{-\frac{n}{m}}]\leq\expect(S)
 \end{equation}
 Because $S$ is the function of the placements of the $n$ balls, it satisfies the Lipschitz condition, we may apply Azuma's inequality to the Doob Martingale and obtain the probability inequality:
$\prob(|S-\expect(S)|\geq \varepsilon m)\leq 2e^{-\varepsilon^2m/2}$.
\end{IEEEproof}
\begin{theorem}
The approximation of the  algorithm~\ref{alg:online duty cycling algorithm} is  $1-e^{-\probeh^2}$, where $\probeh$ is the probability able to harvest energy.
\end{theorem}
\begin{IEEEproof}
Algorithm~\ref{alg:online duty cycling algorithm} obtains  the edge set $\edge$. Let $\edge=\setc_n+\setd_n$, where the set $\setc_n$ contains the synchronous edges and the set $\setd_n$ contains the asynchronous edges.
Notice that each device can harvest energy with probability $\probeh$ so the device and its neighbor can harvest energy in expected  $n=\probeh m$ time slots, \ie $n$ vertexes, where $m=|\period|$.
According to Property~\ref{prop: one deterministic edge}, each vertex can only connect with one vertex of the neighboring device.
In each time slot, the device has probability $n/m$, \ie, $\probeh$, to connect to the $n$ vertexes of its neighbor, and probability $1/m$ to connect to one of  them.
In Algorithm~\ref{alg:online duty cycling algorithm}, the device finds a single synchronous edge with probability $\probeh$, \ie have $\probeh n$ vertexes in expectation. To bound the total number of edges in the set $\setc_n$, we have a balls-in-bins problem with $n$ balls and $n$ bins out of $m$.
We are interested in lower-bounding the number of bins that have at least one ball.
Applying the result in Fact~\ref{fact: ball bin}, we obtain that $\setc_n\geq [1-e^{-\frac{\probeh n}{m}}]n-\varepsilon m$, where $m=|\period|$ and $n=\probeh m$ with probability $1-e^{-\Omega(n)}$.
Similarly, we can obtain that $\setd_n\geq [1-e^{-\frac{\probeh n}{m}}]n'-\varepsilon m$ with probability $1-e^{-\Omega(n')}$, where $n'=2(1-\probeh)m$.
Thus the weight of the set  $\edge_n$ is $\overlaptime_n(\period)=|\setc_n|+\charge|\setd_n|\geq(1-e^{-\frac{\probeh n}{m}})(n+\charge n')-2\varepsilon m\geq(1-e^{-\frac{\probeh n}{m}})(\probeh^2m+2\charge(1-\probeh)\probeh m)-2\varepsilon m$ since $\probeh\leq 1$.

Let $\edge_{OPT}$ denote the edge set obtained by the optimal algorithm OPT.  $\edge_{OPT}=\setc_{OPT}+\setd_{OPT}$, where the set $\setc_{OPT}$ contains the edges with weight one and the set $\setd_{OPT}$ contains the edges with weight $\charge$.
The device and its neighbor have $m\probeh^2$ times of common chance to harvest energy in expectation, and thus have $m\probeh^2$ edges with weight one in expectation, \ie $\setc_{OPT}=m\probeh^2$.
Similarly, it have expected $2m(1-\probeh)\probeh$ edges with weight $\charge$, \ie $\setd_{OPT}=2m(1-\probeh)\probeh$.
Thus the weight of the set $\edge_{OPT}$ is $\overlaptime_{OPT}(\period)=|\setc_{OPT}|+\charge|\setd_{OPT}|=\probeh^2 m+2\charge (1-\probeh)\probeh m$.

Summarize the above analysis, we can obtain that overall \cat, \ie the overall edge weight, obtained by Algorithm~\ref{alg:online duty cycling algorithm} is $\overlaptime_n(\period)/\overlaptime_{OPT}(\period)= (|\setc_n|+\charge|\setd_n|)/(|\setc_{OPT}|+\charge|\setd_{OPT}|)\geq 1-e^{-\frac{\probeh n}{m}}-2\varepsilon m/(|\setc_{OPT}|+\charge|\setd_{OPT}|)$.
We can take the value of  $\varepsilon $ to be sufficient small, and then the approximation ratio of Algorithm~\ref{alg:online duty cycling algorithm} is $1-e^{-\frac{\probeh n}{m}}$, where $\frac{n}{m}$ goes to $\probeh$.
\end{IEEEproof}

In Algorithm~\ref{alg:online duty cycling algorithm}, the harvested energy arrives randomly and the device makes online decision in each time slot. There are $|\period|$ slots totally. In each time slot, the device has $|\period|\probeh^2$ times to find the synchronous edges in expectation, and $|\period|(1-\probeh)\probeh$ times to find the asynchronous edges.
When the device cannot find the synchronous edge, it looks for the asynchronous edges. It takes at most $|\period|\probeh$ steps in expectation to find the asynchronous edge since $(1-\probeh)\probeh<\probeh$, and thus at most $|\period|^2\probeh$ steps overall. Since $\probeh\leq 1$, we have:
\begin{theorem}
The time complexity of Algorithm~\ref{alg:online duty cycling algorithm} is $O(|\period|^2)$.
\end{theorem}
\section{Related Work}
\label{sec:related}
This section reviews the related works on the energy harvesting networks and the duty cycling  technique.

\textbf{Energy harvesting Network}
Tiny energy harvesting devices and networks have been widely researched in recent years.
Liu~\etal designed the novel communication mechanism, ambient backscatter, that enables devices to communicate by backscattering ambient RF with the harvested energy from harvesting the energy from TV signals~\cite{liu2013ambient}.
Xiang~\etal designed a self-sustaining sensing system called Trinity. Trinity harvests energy from the airflow produced
by HVAC outlets to power the sensor nodes, which are then able to communicate with each other~\cite{Li2013Powering}.
Gummeson~\etal designed a ring form-factor wearable device for gesture input, harvesting energy from an NFC-enabled
phone held in a user’s hand~\cite{gummeson2014energy}.
There are other  typical systems, such as Prometheus~\cite{jiang2005perpetual} and Helimote~\cite{hsu2005energy}, able to harvest solar,
         vibration and wind energy~\cite{liu2011perpetual}\cite{ozel2012achieving}.
Among the existing systems, some devices have no storage components and thus use the harvested energy directly~\cite{Huang2013Utility}\cite{he2012energy}.
Some devices store the harvested energy into their batteries while others take the capacitor as the primary buffer and the battery as the second buffer~\cite{jiang2005perpetual}.
The battery suffers from low charging efficiency and long charging duration,  the capacitor has high leakage~\cite{zhu2009leakage}.
For example, the 2000F ultra-capacitor has the high leakage rate up to 43.8\% during the first month~\cite{zhu2009leakage}.

These devices and networks can harvest small amount of  energy from the natural resources continually, which  is not enough to support full duty cycle.
The harvested energy among devices or the networks is heterogenous because of the spatiotemporal variation of the natural resources~\cite{Buchli2014Dynamic}\cite{Li2013Powering}.
Frequent charging/discharging suffers from much energy loss and harms the battery working life.

\textbf{Duty cycling.}
In the energy harvesting devices and networks,  the duty cycling technique is still necessary  because of the limited hardware and insufficient intensity of the natural resources~\cite{jiang2005perpetual}\cite{gu2009esc}.
Previous works on duty cycling can be classified into two groups: deterministic and stochastic duty cycling.
The deterministic duty cycling calculates the amount of active time in each period, such as the adaptive duty cycling~\cite{vigorito2007adaptive}\cite{Buchli2014Dynamic}.
A few works considers not only the amount of the active time but also the moments to wake up, \ie stochastic duty cycling~\cite{Hsin2006Randomly}\cite{ghidini2011energy}.
However, they did not consider that the energy harvesting is a random process~\cite{ho2010markovian}, and the energy profile of each sensor node is time-varying and different among the devices~\cite{kansal2004performance}\cite{zhu2009leakage}.
Furthermore, these previous works adjust  duty cycle only according to the predicted amount of the harvested energy in a duration, and do not consider the moments to be active since the low charging efficiency has great impact on the energy efficiency.

Different from the previous works, this paper addresses the stochastic duty cycling by considering not only the amount of the active time but also the moments to be active over a period.
We consider the spatiotemporal dynamic of the harvested energy and our scheme ensures that each sensor node adjusts its duty cycle by considering both the dynamic harvested energy and low charging efficiency.
\section{Experimental Evaluation}
\label{sec:experiment}
\subsection{Experiment Setting}
\begin{figure}[ht]\centering
\subfigure[Solar sensor node]{\label{subfig:experiment seting one}\includegraphics[scale=.23,bb=35 480 559 806]{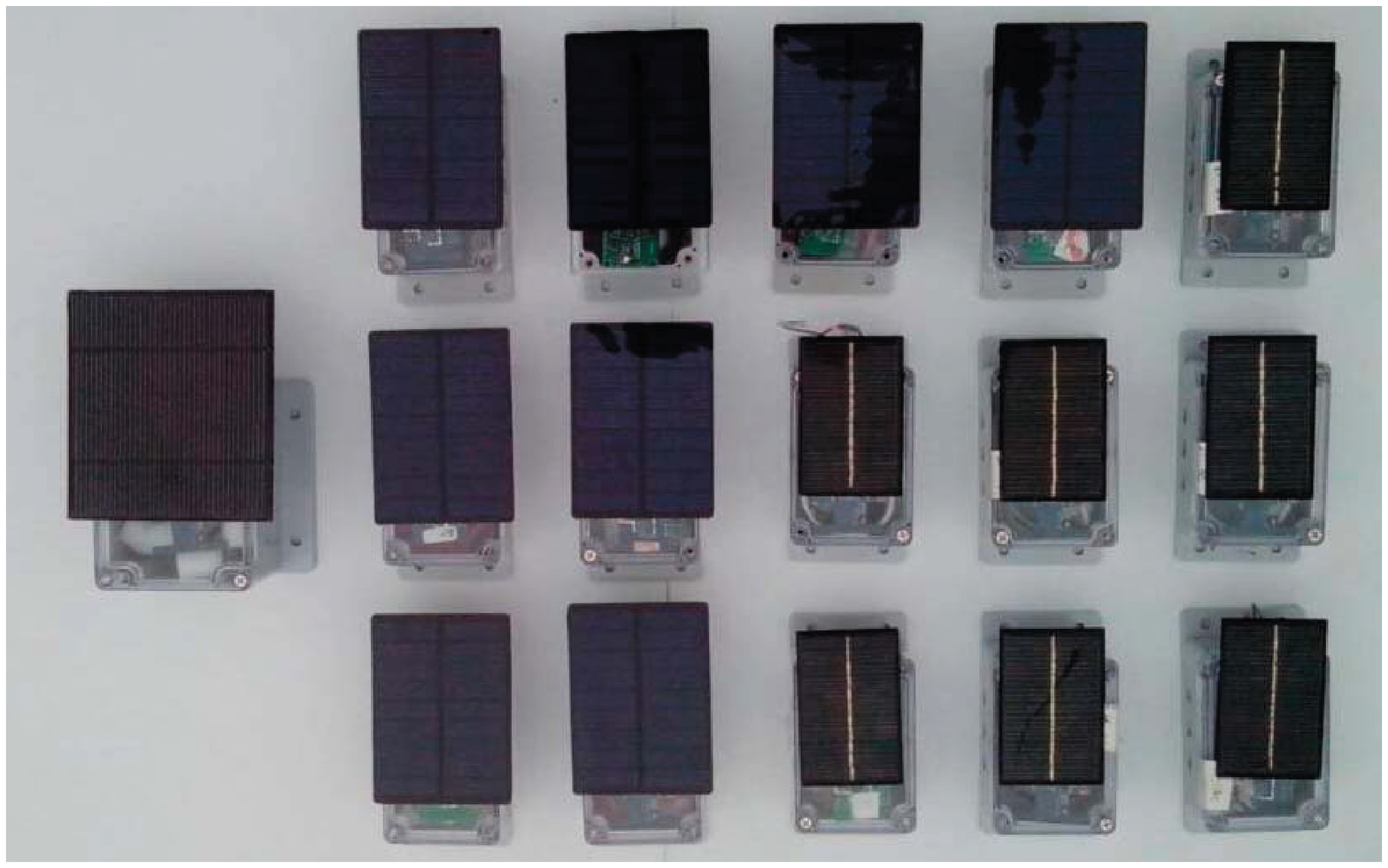}}
\hspace{0.5cm}
\subfigure[Sensor deployment]{\label{subfig:experiment seting two}\includegraphics[scale=.51, bb=220 502 400 652]{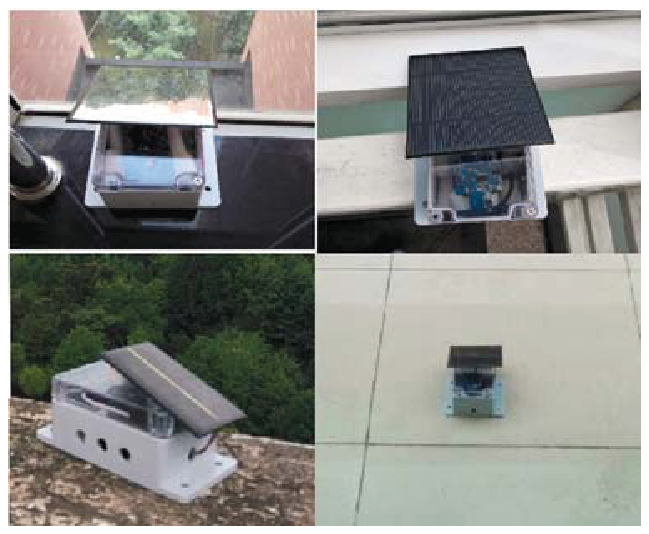}}
\caption{\label{fig:solar experiment} Experiment devices and deployment.}
\end{figure}
This section conducts the experiment to evaluate the performance of the offline/online algorithms.
The experiment device is the solar sensor nodes in Figure~\ref{subfig:experiment seting one} and is able to harvest the solar energy and communicate with others~\cite{shen2009solarmote}.
The controller in the node can shift the harvested energy to charge the battery or to support its operation.
The solar sensor nodes are deployed around our lab building in Figure~\ref{subfig:experiment seting two} and have different time to access the solar energy.
In this experiment, we consider the communication among node pairs instead of networked communication,  such as data aggregation~\cite{wang2013near} or packet deliver~\cite{dong2014measurement}.
Fifteen nodes and one sink are deployed to harvest solar energy, and to sample and deliver the data of the temperature and luminous density with the harvested energy. There are totally eight pair of devices, among which seven pair of devices runs the offline and online  algorithms. The sink connects with a personal computer, which collects data from the devices.
The period and the slot are respectively set to be 10 hours and 1 minute.
\subsection{Result Analysis}
\begin{figure}[h]\centering
\subfigure[CAT percentage]{\label{subfig: cat with probability}\includegraphics[scale=.45,bb=65 581 224 723]{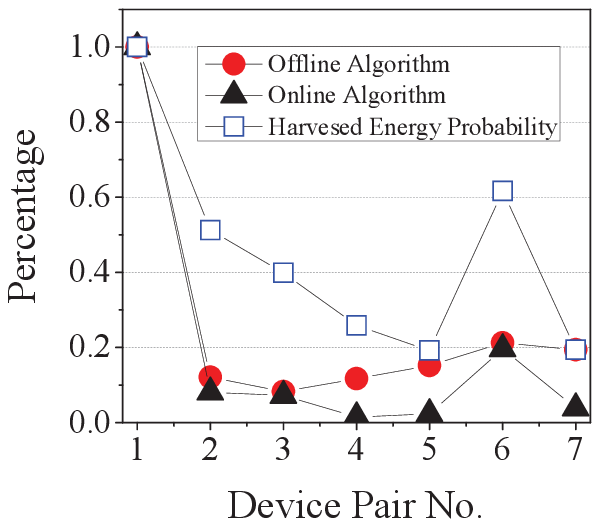}}
\hspace{0.1cm}
\subfigure[SAT percentage]{\label{subfig:sat with probability}\includegraphics[scale=.45, bb=65 581 224 723]{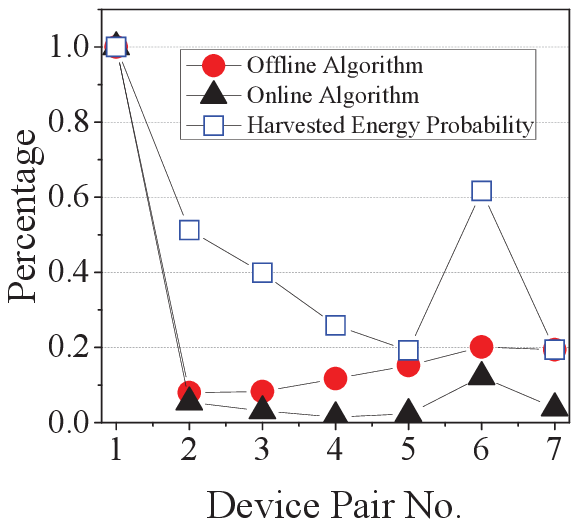}}
\hspace{0.1cm}
\subfigure[Comparison over two algorithms]{\label{subfig:cat and sat percentage}\includegraphics[scale=.45, bb=65 581 224 723]{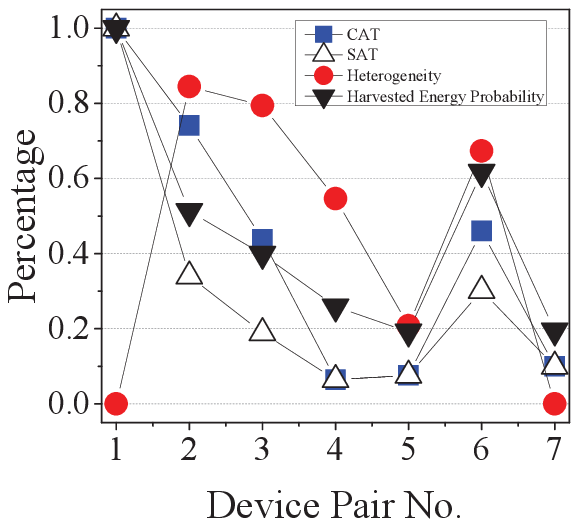}}
\caption{\label{fig: percentage of performance} Two parameters, $\probeh$ and heterogeneity, impact the algorithm performance.}
\end{figure}

We first measure the metrics: \cat and Synchronous Active Time (SAT), and then analyze the factors impact the metrics.
 SAT is the sum of the synchronous edges' weight.
Figure~\ref{subfig: one node one day} shows the \cat obtained by the offline and online  algorithms among the seven different pairs of devices.
The \cat obtained by the offline algorithm ranges from 49.45   to 600 minutes while it ranges from 9.04 to 600 minutes under the online algorithm.
Figure \ref{subfig: cat with probability} shows the percentage of the \cat over the period length respectively under the offline and online algorithms.
\cat can be from 8.24\% to 100\% of the period under the offline algorithm, and from 2.30\% to 100\% of the period length under the online algorithm.
Figure~\ref{subfig:one node two day} shows that SAT is from 47.77 to 600 minutes under the offline algorithm, and from 9.04   to 600 minutes under the online algorithm.
Figure \ref{subfig:sat with probability} shows the percentage of the SAT over the period length respectively under the offline and online algorithms.
SAT is from 7.96\% to 100\% of the period under the offline algorithm, and from 1.52\% to 100\% of the period under the online algorithm.
Figure~\ref{subfig:cat and sat percentage} shows the performance of the online algorithm comparison to the offline algorithm.
The \cat under the online algorithm can be from 10\% to 100\%, and 52.68\% on average  of that under the offline algorithm.
The SAT under the online algorithm can be from 6.4\% to 100\%, and 27.45\% on average  of that under the offline algorithm.
\begin{figure}[h]\centering
\subfigure[CAT]{\label{subfig: one node one day}\includegraphics[scale=.45,bb=62 575 301 725]{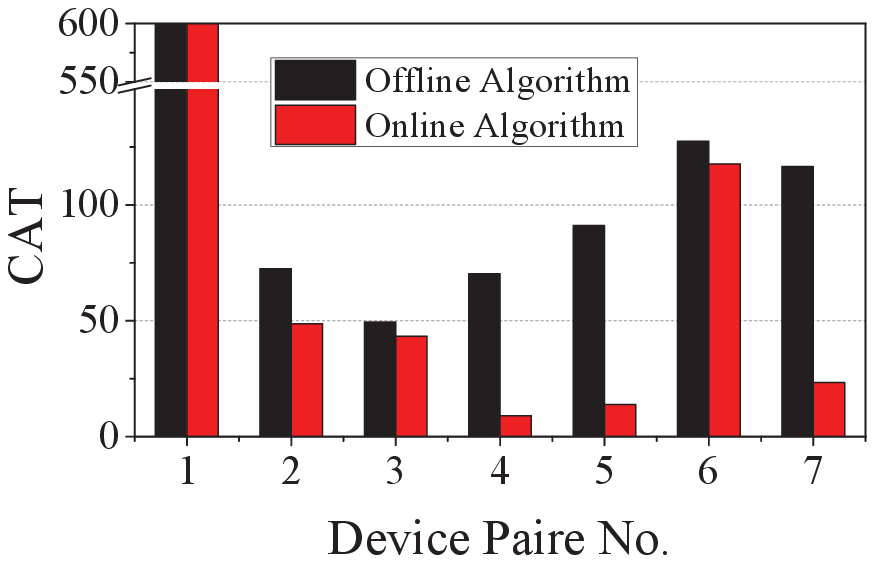}}
\hspace{0.2cm}
\subfigure[SAT]{\label{subfig:one node two day}\includegraphics[scale=.45, bb=62 575 301 725]{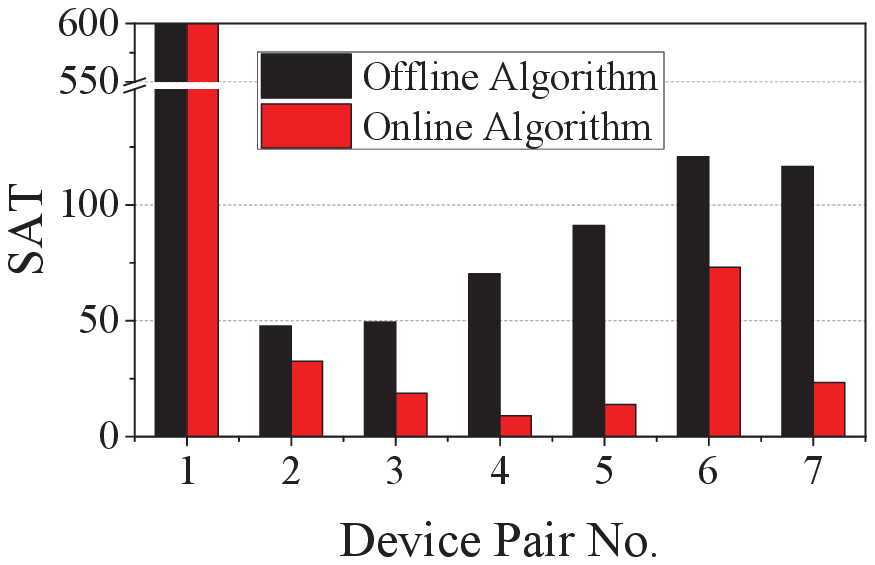}}
\caption{\label{fig: common active experiment} The energy harvested by a same device in three days.}
\end{figure}

The experimental results show that   the \cat and SAT are quite different among the different pairs of devices.
We find that two parameters can describe the impact: the harvested energy probability, \ie $\probeh$, and the heterogeneity.
The heterogeneity is defined as the percentage of the non-\cat between a pair of devices. Higher heterogeneity results in less \cat among devices.
Figure~\ref{subfig: cat with probability} and \ref{subfig:sat with probability} show that $\probeh$ is highly in accordance with the percentage of \cat and SAT.
Figure~\ref{subfig:cat and sat percentage} also show that  $\probeh$  is highly in accordance with the relative performance of the online algorithm in comparison with the offline algorithm.
When $\probeh$ closes to 1, \cat obtained by the online algorithm can close to the offline algorithm.
Observe the case: No.1 pair of device,  the performance of the online algorithm is same with the offline algorithm when $\probeh=1$.
This figure also shows that the metric: heterogeneity, is a good measure for the performance difference between the offline and online algorithms.
The lower the heterogeneity is, the performance of the online algorithm is closer to the offline algorithm.
\section{Conclusion}
\label{sec:conclusion}
This paper investigates the stochastic duty cycling problem under the random  harvested energy and low charging efficiency.
Two cases: offline and online duty cycling, are studied. For the offline case, this paper designs the offline duty cycling algorithm with the technique of the maximal weighted bipartite matching.
For the online case, this paper designs the online duty cycling algorithm with the approximation ratio of $1-e^{-\probeh^2}$.
This paper also conducts the experiment to evaluate the performance of the offline and online duty cycling algorithms.


\bibliographystyle{unsrt}
\bibliography{pakeyref}
\end{document}